\documentclass[aps,prl,superscriptaddress,showpacs,twocolumn]{revtex4}
\usepackage{epsfig}
\usepackage{color}

\newcommand{\Nabla}{{\mbox{\boldmath$\nabla$}}}

\begin{document}

\title{Disorder-Induced Resistive Anomaly Near Ferromagnetic Phase
Transitions}

\author{Carsten Timm}
\affiliation{Institut f\"ur Theoretische Physik, Freie Universit\"at Berlin, Arnimallee
14, 14195 Berlin, Germany}
\author{M. E.\ Raikh}
\affiliation{Department of Physics, University of Utah, Salt Lake City, 
UT 84112}
\author{Felix von Oppen}
\affiliation{Institut f\"ur Theoretische Physik, Freie Universit\"at Berlin, Arnimallee
14, 14195 Berlin, Germany}

\date{August 27, 2004}

\begin{abstract}
We show that the resistivity $\rho(T)$ of disordered ferromagnets near, and above, the Curie temperature 
$T_c$ generically exhibits a stronger anomaly than the scaling-based Fisher-Langer prediction.
Treating transport beyond the Boltzmann description, we find that within mean-field theory,
$d\rho/dT$ exhibits a $|T-T_c|^{-1/2}$-singularity near $T_c$.    
Our results, being solely due to impurities, are  
relevant to ferromagnets with low $T_c$, such as SrRuO$_3$ or diluted magnetic semiconductors,
whose mobility near $T_c$ is limited by disorder.
\end{abstract}


\maketitle

{\it Introduction.}---It was first observed by Gerlach \cite{Gerlach} that the resistivity of itinerant 
ferromagnets exhibits an anomalous temperature dependence in the vicinity of the Curie temperature $T_c$.
This feature was later reproduced with much higher experimental accuracy by Craig {\it et al.} 
\cite{Craig} as well as others \cite{other}.
Dating back to the seminal works of de Gennes and Friedel \cite{deGennes} as well as Fisher and Langer
\cite{Fisher}, this resistive anomaly is conventionally explained in terms of {\it coherent} scattering 
of carriers by large blocks of spins whose size is determined by the magnetic correlation length $\xi(T)$.
As the temperature approaches $T_c$, $\xi(T)$ diverges, making the scattering more efficient. 
 
de Gennes and Friedel \cite{deGennes} studied the resistive anomaly of ferromagnets within mean-field 
(MF) theory. They argue that due to critical slowing down, spin fluctuations can be treated as 
effectively static. The ensuing result for the coherent transport 
scattering rate from spin fluctuations above $T_c$ is 
succinctly summarized by \cite{deGennes}
\begin{equation}
{\tau_0\over \tau} = {1\over 4} \int dx\, {x^3 \over t + (k_F a x)^2}.
\label{deGennes}
\end{equation}
Here, the scattering rate is normalized to the rate $1/\tau_0$ for incoherent scattering. The 
denominator of the integrand arises from the conventional MF (Ornstein-Zernike) correlator for spin fluctuations,
with $t=(T-T_c)/T_c$ denoting the reduced temperature. $a$ is a microscopic length of
the order of the lattice constant. The integration variable $x$ denotes 
the transferred momentum in units of the Fermi wave vector $k_F$. The numerator incorporates
the usual factor $1-\cos\theta$ (with $\theta$ the scattering angle) in the transport scattering rate. 
The integral in Eq.\ (\ref{deGennes}) yields 
$\tau_0/\tau(t)=\tau_0/\tau(0)+(1/8(k_Fa)^4)\, t\ln t $,
implying a singularity of the resistivity of the form
$\rho(T) = \rho_0 - b t \ln (1/t)$ with $b>0$ when approaching $T_c$ from above. 

Fisher and Langer \cite{Fisher} noticed that this singularity emerges from the lower limit 
of the integral in Eq.\ (\ref{deGennes}) while the body of the integrand is dominated 
by large wave vectors. Within MF theory, the large-wave-vector behavior of
the spin-spin correlator is non-singular. The central assertion of Ref.\ \cite{Fisher} is that 
there exists a \emph{singular} contribution to this correlator
at large wave vectors when going beyond the MF approximation, with the singularity governed by 
anomalous dimensions. Fisher and Langer conclude that while below $T_c$, the predictions of 
Ref.\ \cite{deGennes} are essentially correct, there is {\it no singularity within 
MF approximation} when approaching $T_c$ from above. 

This conclusion rests on their important physical observation that Eq.\ (\ref{deGennes}) becomes
inapplicable for $t\ll (\ell/a)^2$, or equivalently, when the correlation length $\xi(T)$ exceeds 
the mean free path $\ell$ (due to phonon or impurity scattering). The reason is that the carriers
can be viewed as plane waves only over distances shorter than $\ell$ and thus, they
are no longer susceptible to the order in the spin configuration beyond $\ell$. 

Later, the ideas of Refs.\ \cite{deGennes,Fisher} were extended to include realistic features
of ferromagnets \cite{geldart77} and to the critical behavior of other quantities 
such as the spin-flip scattering rate \cite{ora75,geldart85,schuller78}. 
The effect of a finite mean free path on the resistivity was studied by a
number of
approaches, ranging from replacing the $\delta$-function in the golden rule by 
a Lorentzian \cite{geldart85} to smearing the Ornstein-Zernike correlator in order to eliminate the pole 
\cite{Kataoka}.

{\it All} of these approaches 
are based on the Boltzmann-equation 
formalism. The prime message of this paper is that in the small-$t$ limit, when the correlation 
length $\xi(T)$ exceeds the mean free path $\ell$, the Boltzmann approach {\it fails}.
The reason for this is that for $\xi(T)\gg \ell$, the smooth variations of the magnetization 
(on the scale $\xi(T)$) are ``explored" by {\it diffusing} carriers. By contrast, the Boltzmann 
approach prescribes to treat scattering from both short-range impurities and the smooth variations
of the magnetization
on equal footing, i.e., to add their partial scattering rates. 

The consequences of going beyond the Boltzmann approach are drastic. In fact,
as demonstrated below, instead of smearing the resistive anomaly, impurities
cause a much stronger singularity, even within MF theory. Quantitatively,
we find $d^2\rho/dT^2\sim t^{-3/2}$ sufficiently close to $T_c$,
as opposed to $d^2\rho/dT^2\sim t^{-1}$ for $\xi(T) \ll \ell$ \cite{deGennes}.

\begin{figure}\centering
\epsfxsize6cm\epsfbox{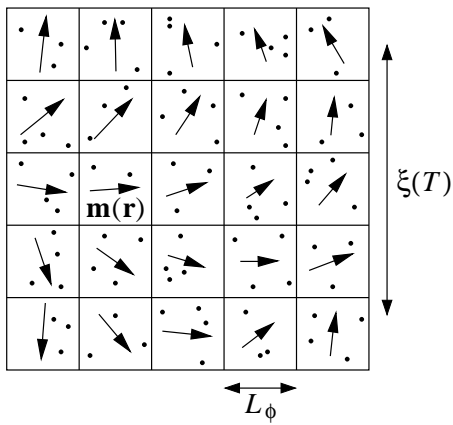} \vskip .3truecm 
\caption{Schematic representation of resistor network describing disordered
ferromagnets when the correlation length $\xi(T)$ exceeds the phase-coherence length $L_\phi$.
Each block of size $L_\phi$ constitutes a resistor with coarse-grained magnetization (arrows)
and random impurities (dots). 
} \label{fig1}
\end{figure}

Our reasoning goes as follows. 
The adequate description of electric transport on scales larger
than the phase-breaking length $L_\phi$ is a network of resistors, made up of cubes of size $L_\phi$,
as illustrated in Fig.\ \ref{fig1}. This network is inhomogeneous due to the different spin 
and disorder configurations in each cube. Contrary to the Boltzmann prescription, it is essential to first 
compute the effective resistivity of the inhomogeneous network and to perform the disorder and 
thermal averages only as the last step of the calculation. Then, correlations between 
distant spins affect the resistance through the inhomogeneous current and field distribution over 
the network, leading to a true singularity for $t\to 0$ even within MF theory.

To illustrate the importance of performing the impurity and thermal averages at the last step of the 
calculation, consider a minimal model of two 
macroscopic resistors in sequence. 
These resistors differ in both impurity and spin configurations. 
Within the Boltzmann approach, both resistances are equal to $1/\overline{\sigma}$ upon 
configurational and thermal averaging, yielding a total resistance 
of $2/\overline{\sigma}$. However, for the {\it actual} distribution of impurities 
and spins, their conductivities differ, so that $\sigma_1 = \overline{\sigma}+\delta\sigma/2$ while
$\sigma_2 = \overline{\sigma}-\delta\sigma/2$. Then, the effective resistance becomes 
$(\sigma_1+\sigma_2)/\sigma_1\sigma_2 \simeq 2/\overline{\sigma} + \delta\sigma^2 /2\overline{\sigma}^3$
(assuming $\delta\sigma \ll \overline{\sigma}$). This involves an {\it additional} term
$\delta\sigma^2 /2\overline{\sigma}^3$ with non-zero average.

{\it Effective conductivity of an inhomogeneous medium.}---Remarkably, the effective resistivity $\rho_{\rm eff}
=1/\sigma_{\rm eff}$ can be computed for an arbitrary realization $\sigma({\bf r})$
of the local conductivity, provided that the relative variation in $\sigma({\bf r})$ is weak \cite{Landau}. 
To see this, we decompose the current, the conductivity, and the electric field into averages ${\bf j}_0$, 
$\sigma_0$,
and ${\bf E}_0$ and spatially fluctuating contributions $\delta{\bf j}({\bf r})$, 
$\delta\sigma({\bf r})$, and $\delta{\bf E}({\bf r})$. From Ohm's law, we have
\begin{eqnarray}
  {\bf j}_0 &=& \sigma_0 {\bf E}_0 + \langle \delta\sigma({\bf r})\delta{\bf E}({\bf r})\rangle, 
  \label{aver}\\
  \delta{\bf j}({\bf r}) &=& \delta \sigma({\bf r}){\bf E}_0 + \sigma_0 \delta {\bf E}({\bf r}).
  \label{fluc}
\end{eqnarray}
Here, the brackets denote a spatial average. Combining the continuity equation $\Nabla\cdot\delta{\bf j}=0$
and Maxwell's equation $\Nabla\times\delta {\bf E} = 0$ with Eq.\ (\ref{fluc}), one obtains
$\delta {\bf E}({\bf q}) = -{\hat{\bf q}}\, {\hat{\bf q}}\cdot {\bf E} {\delta\sigma({\bf q})/\sigma_0}$,
where ${\hat {\bf q}}$ denotes the unit vector in the direction of the wave vector ${\bf q}$. Inserting this
into Eq.\ (\ref{aver}), one finds for the effective macroscopic conductivity $\sigma_{\rm eff}$ (defined by
${\bf j}_0 = \sigma_{\rm eff} {\bf E}_0$) in three dimensions
\begin{equation}
   \sigma_{\rm eff} = \sigma_0 - {\langle [\delta\sigma({\bf r})]^2\rangle \over
      3\sigma_0}.
      \label{effcond}
\end{equation}
Remarkably, this result is independent of the geometry of the conductivity variations. 
To illustrate this, consider a sample with a 50-50 random mixture of 
domains of conductivities $\sigma_1$ and $\sigma_2$, where $|\sigma_1-\sigma_2|\ll \sigma_1,
\sigma_2$. In this case, Eq.\ (\ref{effcond}) yields $\sigma_{\rm eff} = \sigma_0 - (\sigma_1-\sigma_2)^2/
12\sigma_0$ 
, which is 
completely independent of the arrangement of domains. It is interesting to remark that in two 
dimensions, one can find an exact, geometry-independent 
expression for such two-phase systems which is valid for arbitrarily 
large inhomogeneities $|\sigma_2-\sigma_1|/(\sigma_1+\sigma_2)$ \cite{Dykhne}. 

{\it Effective conductivity due to spin fluctuations.}---The conductance of a phase-coherent sample exhibits 
random, sample-specific, and reproducible variations as a function of external parameters such as the Fermi 
energy $E_F$. These conductance fluctuations arise due to interference between different 
elastic scattering paths of a carrier that diffuses through the sample \cite{UCF}. 
Thus, the conductance $g({\bf r},E_F)$ varies from block to block because of their different impurity
configurations. The fluctuations in the conductivity entering Eq.\ (\ref{effcond}) can then be 
expressed as 
\begin{equation}
  \delta\sigma({\bf r}) = \delta g({\bf r},E_F;{\bf m}({\bf r}))/L_\phi .
  \label{general}
\end{equation}
{We assume that the system is so large that domains with any magnetization
combined with any disorder realization appear. Using this \emph{ergodicity
assumption} we replace the \emph{spatial} average in Eq.~(\ref{effcond}) by
independent disorder and thermal (magnetization) averages.}

\begin{figure}\centering
\epsfxsize7cm\epsfbox{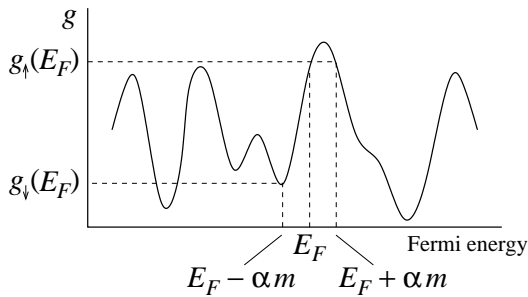} \vskip .3truecm 
\caption{Sample-specific variation of $g(E_F)$ for a block of the resistor network (UCF) 
in the absence of spin. The conductances for each spin direction 
are obtained by including equal, but opposite exchange-induced Zeeman shifts of the Fermi 
energy. As indicated, this leads to a difference in the conductances for the two spin directions.
} \label{fig3}
\end{figure}

{\it Two spin subbands.}---In 
our case, the external para\-meter is a vector, namely the magnetization ${\bf m}({\bf r})$. 
To proceed, we first consider the simplest case in which the dominant effect of the
impurity spins on the carriers is an effective Zeeman field arising from the exchange interaction, which is 
proportional to the magnetization $m({\bf r})$. Then the {\it coarse-grained} magnetization 
${\bf m}({\bf r})$, which is averaged 
over each cube of size $L_\phi$, can be incorporated via equal, but opposite energy 
shifts $\pm \alpha m({\bf r})$ for spin-up and spin-down carriers, and $\delta\sigma$ becomes
a sum of contributions from the two spin projections,
\begin{eqnarray}
   \delta\sigma ({\bf r}) &=& [\delta g_\uparrow({\bf r},E_F) + \delta g_\downarrow({\bf r},E_F)]/L_\phi
    \nonumber\\
    &=&[\delta g({\bf r},E_F^+({\bf r})) + \delta g({\bf r},E_F^-({\bf r}))] /L_\phi.
\label{delsig}
\end{eqnarray}
Here, $g({\bf r},E_F)$ is the {\it spinless} conductance of the cube at position ${\bf r}$ and we
have introduced the shifted Fermi energies
$E_F^\pm = E_F \pm \alpha m$, see Fig.\ \ref{fig3}. We used that carriers with both spin projections 
are scattered from the {\it same} impurities.  
Substituting Eq.\ (\ref{delsig}) into Eq.\ (\ref{effcond}), we obtain
\begin{eqnarray}
\rho_{\rm eff}\!&\!-\!&\!\rho_0 = {2\rho_0^3\over 3L_\phi^2}
\left\{ 
\langle [\delta g ({\bf r},E_F^+({\bf r}))]^2 \rangle
+\langle [\delta g ({\bf r},E_F^-({\bf r}))]^2 \rangle
\right. 
\nonumber\\
&& \left.+ 2\, \langle \delta g ({\bf r},E_F+\alpha m({\bf r}))\,
\delta g({\bf r},E_F- \alpha m({\bf r})) \rangle
\right\} .
\label{correlator}
\end{eqnarray} 
{The first two terms on the rhs describe universal conductance
fluctuations, which are independent of the Fermi energy and thus
of $m({\bf r})$ \cite{UCF}. Therefore, the only $t$-dependence comes 
from the conductance correlator.}

It is well known \cite{UCF} that this conductance correlator
is a function $F(x)$ of the dimensionless ratio $x=\alpha m/E_c$, where 
$E_c=D/L^2_\phi$ is the inverse diffusion time through a block of size $L_\phi$ 
($D$ denotes the diffusion constant). 
In three dimensions, the asymptotic behaviors of $F(x)$ are given by \cite{UCF}
$F(x) =  F(0) (1 - C_1 x^2) $ for $ x\ll 1 $
and  $F(x) =  F(0) C_2 x^{-1/2} $ for $ x \gg 1 $,
where $C_1$ and $C_2$ are constants of order unity. The remaining step is to substitute $F(x)$ into
Eq.\ ({\ref{correlator}}) and to perform the thermodynamic average over ${\bf m}$, using the 
MF distribution for $t>0$,
\begin{equation}
  {\cal P}[{\bf m}({\bf r})] 
  \propto \exp \left\{ -{c\over 2T} \int d^3r \left[a^2\, \nabla_i {\bf m}\cdot\nabla_i {\bf m}  
  + t\, {\bf m}^2\right]
  \right\}.
  \label{Landau}
\end{equation}
Here, $ca^2$ is a spin stiffness. 
Using this distribution, one finds for the magnetization
fluctuations 
\begin{equation}
 \langle {\bf m}^2({\bf r})\rangle  = \int_{q<1/L_\phi} {d^3q\over (2\pi)^3}\,
 {3T_c/c\over t + q^2 a^2 }.
\label{msq}
\end{equation}
The restriction in the range of the ${\bf q}$ integration accounts for the fact that we are computing
fluctuations of the coarse-grained magnetization. The appearance of $t$ in the denominator
of Eq.\ (\ref{msq}) manifests the fact that the {\it net} strength of fluctuations grows when 
approaching $T_c$. This is in contrast to the conventional origin of the $t$-dependence, namely the spin
{\it correlations}. Performing the integration in Eq.\ (\ref{msq}), we obtain
$\langle {\bf m}^2({\bf r})\rangle = (3T_c/2\pi^2 c a^2) \left[ 1/L_\phi - (1 / \xi(T))\arctan (\xi(T)/L_\phi)
\right]$ in terms of the MF correlation length $\xi(T) = a/\sqrt{t}$. 

As the next step, we replace $m({\bf r})$ in Eq.\ (\ref{correlator}) by $ \langle {\bf m}^2({\bf r})\rangle^{1/2}$.
Using the distribution
(\ref{Landau}), this procedure can be shown to be exact in the limits of small and large $x$. Expanding the 
result in the small parameter $L_\phi/\xi(T)= L_\phi \sqrt{t}/ a$, we readily obtain
\begin{eqnarray}
    \rho_{\rm eff}-\rho_0 =
     {2\rho_0^3 
    \over 3L_\phi^2}\left[  F(x_0) - F^\prime(x_0) x_0 {\pi L_\phi \sqrt{t}\over 4a} \right],
        \label{final1}
\end{eqnarray}
where $x_0=(\alpha/E_c)(3T_c/2\pi^2 c a^2 L_\phi)^{1/2}$.  The $t$-dependent part of $\rho_{\rm eff}$
is given by the second term on the rhs \cite{note.critical}.

So far, our model completely disregards 
spin-orbit (SO) coupling which is present in the vast majority of ferromagnets. 
Below, we incorporate SO coupling into the calculation of the effective resistivity.

{\it Spin-orbit coupling.}---In the presence of SO coupling, the variance 
$\langle [\delta g({\bf r}, E_F; {\bf m}({\bf r}))]^2\rangle_{\rm imp}$ 
increases by a factor of two when applying a sufficiently strong Zeeman
field \cite{Beenakker}. In terms of the relevant dimensionless measure of the 
exchange-induced Zeeman field
$y=\alpha m\tau_{\rm so}$ (here $\tau_{\rm so}$ is the SO time), 
we can then write $\langle [\delta g({\bf r}, E_F; {\bf m}({\bf r}))]^2\rangle_{\rm imp} = H(y)$ with 
$H(\infty)/H(0) = 2$ \cite{Beenakker}. Combining Eqs.\ (\ref{general}) and (\ref{effcond}), substituting $H(y)$,
and proceeding as in the derivation of Eq.\ (\ref{final1}), we obtain
\cite{note.critical}
\begin{eqnarray}
    \rho_{\rm eff}- \rho_0 =
    {2 \rho_0^3 
    \over 3L_\phi^2}\left[  H(y_0) - H^\prime(y_0) y_0 {\pi L_\phi \sqrt{t}\over 4a} \right],
        \label{final2}
\end{eqnarray}
where $y_0=\alpha \tau_{\rm so} (3T_c /2\pi^2 c a^2 L_\phi)^{1/2}$. Eq.\ (\ref{final2})  
assumes that the SO length $\ell_{\rm so}=(D\tau_{\rm so})^{1/2}$ is smaller 
than $L_\phi$.

\begin{figure}\centering
\epsfxsize7cm\epsfbox{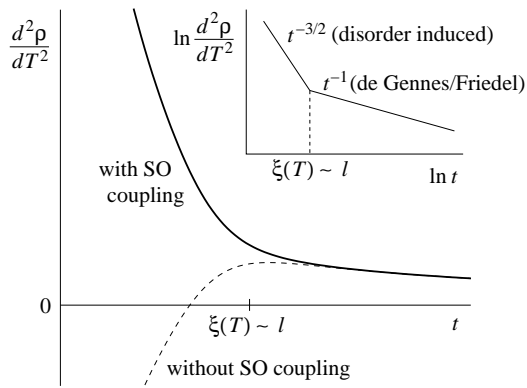} \vskip .3truecm 
\caption{Resistive anomaly within MF theory for $L_\phi\sim \ell$ (schematic)
with (full line) and without (dashed line) SO coupling. 
The anomaly is described by the de Gennes-Friedel mechanism for 
$\xi(T) \ll \ell$, while the disorder-induced mechanism of this paper dominates
closer to $T_c$ where $\xi(T) \gg L_\phi$. When $L_\phi \gg \ell$, there is an additional 
intermediate regime. Inset: anomaly with SO coupling in a log-log plot. 
} \label{fig2}
\end{figure}

{\it Discussion.}---By going beyond the Boltzmann description in calculating the critical behavior
of the resistivity, we have expressed $\rho_{\rm eff}$ through mesoscopic characteristics. 
Recall that our results, Eqs.\ (\ref{final1}) and (\ref{final2}), were obtained within MF
theory so that the singularity in the $t$-dependence of $\rho_0$ on the lhs 
of these equations, arising within the Boltzmann formalism,  is suppressed 
by impurities, since $\xi(T) \gg \ell$. Eqs.\ (\ref{final1}) and (\ref{final2})  show that in addition to this suppression, impurity scattering leads to a $\sqrt{t}$ singularity which is much stronger than 
the de Gennes-Friedel result $t\ln (1/t)$ and the large-wave-vector Fisher-Langer contribution.
This singularity, which in essence is governed by Kirchhoff's laws, constitutes our central result. 
To resolve the anomaly on top of a monotonous phonon contribution to $\rho$, it is customary to
consider $d^2\rho /dT^2$. Then, our disorder-induced MF
anomaly becomes $d^2\rho /dT^2 \sim t^{-3/2}$. 
In a log-log plot, the slope is $-1$ in the Boltzmann regime 
$\xi(T) \ll \ell$ and $-3/2$ for the disorder-induced anomaly,
as shown in Fig. \ref{fig2}.

According to Eqs.\ (\ref{final1}) and (\ref{final2}), the sign of the anomaly is governed by the 
sign of either $F^\prime(x_0)$ or $H^\prime(y_0)$, depending on the strength of 
SO coupling. Since $F^\prime(x_0)<0$ while 
$H^\prime(y_0)>0$, the disorder-induced anomaly corresponds to a {\it decrease} of $d^2\rho /dT^2$ when 
approaching $T_c$ from above in the absence of SO coupling and to an {\it increase} in the 
realistic case of strong SO coupling. 
The difference in signs comes about because the Zeeman field suppresses the correlator in Eq.\
(\ref{correlator}), while it increases the 
UCF for strong SO interactions. 

The magnitude of the disorder-induced anomaly is controlled by $F(0)$ and $H(0)$, which are the 
variances of the conductance, $\langle(\delta g)^2\rangle_{\rm imp}$, cf.\  
Eqs.\ (\ref{final1}) and $(\ref{final2})$.   
This quantity assumes different values in different regimes
which are defined by the relations between the relevant lengths, namely $L_\phi$, the thermal length
$L_T=(D/T)^{1/2}$, and the spin-flip length $\ell_s$. 
In the simplest case, when $L_\phi$ is the smallest length, $L_\phi \ll L_T,\ell_s$, 
we have $\langle(\delta g)^2\rangle_{\rm imp}\sim (e^2/h)^2$ \cite{UCF}. If $L_\phi$ is 
larger than $L_T$ or $\ell_s$, then $\langle(\delta g)^2\rangle_{\rm imp}$ is suppressed
below the universal limit, $\langle(\delta g)^2\rangle_{\rm imp}\sim (e^2/h)^2 
[{\rm min}\{L_T,\ell_s\}/ L_\phi]$ \cite{UCF}.

The disorder-induced anomaly proposed in this paper is most relevant 
to ferromagnets with high resistance and low $T_c$ since in such systems (i) the mean free path 
is dominated by impurity scattering \cite{foot2} and (ii) the phase-coherence length exceeds the mean free
path at $T_c$. Natural candidates for disordered low-$T_c$ ferromagnets are SrRuO$_3$ 
\cite{Kapitulnik} which  belongs to the class of poor metals 
\cite{Kivelson} as well as 
diluted magnetic semiconductors (DMS) \cite{Timm,Brey} 
which lately attracted considerable attention in view of possible 
spintronics applications \cite{review}. Indeed,
both types of materials show resistive anomalies which differ significantly from the predictions 
of Fisher-Langer 
theory. Metallic samples of DMS exhibit pronounced maxima near $T_c$ even in $\rho$ vs.\ $T$
\cite{esch97,matsukura98,ohno99,hayashi01,potashnik01,edmonds02,wojtowicz03}, which are unrelated to the $T=0$ metal-insulator transition 
\cite{ohno99,footDMS}.

In closing, it is interesting to point out that our principal result, namely the 
enhancement of the resistive anomaly by disorder, can be viewed in perspective of the enhanced
coupling of the spin fluctuations to the carriers, expected due to the diffusive carrier dynamics.

Two of us (MER and FvO) acknowledge the hospitality of the Weizmann Institute
(supported by the Einstein Center and LSF grant HPRI-CT-2001-00114) 
where this work was initiated, as well as of the MPI for Complex Systems. 
This work was also supported by NSF-DAAD  
Grant No.\ INT-0231010 and Sfb 290.


\begin{thebibliography} {50}
\bibitem{foot2} For $\ell_{\rm imp} \gg \ell$, the disorder-induced anomaly is reduced 
by a reduction in the UCF, $\langle (\delta g)^2 \rangle_{\rm imp}
\sim (e^2/h)^2(\ell/\ell_{\rm imp})$.
\bibitem{footDMS}{A detailed analysis should take into account the small
value of $k_Fa \ll 1$ in DMS. According to Ref.\ \cite{deGennes}, this
implies a wide resistivity peak with  $\delta\rho \sim \rho$. The anomaly
discussed here sits on top of this wide maximum in the  immediate vicinity
of $T_c$.}
\end{thebibliography}
\end{document}